\documentclass[slac_one]{revtex4}
\usepackage{graphicx}
\usepackage{fancyhdr}
\usepackage{amsmath,epsfig}
\newcommand{\msbar}{\ensuremath{\overline{\text{MS}}}}
\newcommand{\xmu}{\ensuremath{x_{\mu}}}
\newcommand{\xmuopt}{\ensuremath{x_{\mu}^\text{opt}}}
\newcommand{\xmuech}{\ensuremath{x_{\mu}^\text{ECH}}}
\newcommand{\as}{\ensuremath{\alpha_s}}
\newcommand{\asZ}{\ensuremath{\alpha_s(M_{\text{Z}})}}
\newcommand{\oassq}{\ensuremath{{\cal{O}}(\alpha_s^2)}}
\newcommand{\epem}{\ensuremath{\text{e}^+\text{e}^-}}
\newcommand{\lambdamsbar}{\ensuremath{\Lambda_{\overline{\text{MS}}}}}
\newcommand{\gev}{{\ifmmode   \mbox{Ge\kern-0.2exV}\else
    Ge\kern-0.2exV\nolinebreak\fi}}
\newcommand{\delphi}{{\sc Delphi}}

\newcommand{\opal}{{\sc Opal}}

\newcommand{\lep}{{\sc Lep}}
\newcommand{\petra}{{\sc Petra}}

\begin{document}

\title{Effective Charges in Practice}

%

\author{Klaus Hamacher}
\affiliation{Fachbereich Mathematik und Naturwissenschaften,\\ Bergische Universit\"at, Gau\ss{}stra\ss{}e 20, 
42097 Wuppertal, Germany,\\ DELPHI Collaboration}

\begin{abstract}
Experimental results on event shapes obtained within (or related to)
the method of Effective Charges are discussed in view of measurements of the
strong  coupling, \as, the $\beta$-function and non-perturbative contributions
to event shapes. The data strongly advocate to use of the ECH scheme
instead of the conventional \msbar\ scheme.
\end{abstract}

\maketitle

\thispagestyle{fancy}


\section{Introductory Remarks}
\subsection{Assorted Theoretical Formulae and Results}
Today the \msbar-scheme is the {\it de facto} standard used for
comparisons of QCD predictions as well as experimental results.
It was introduced for practical reason, as an offspring
of dimensional regularisation, 
 whereas
the method of Effective Charges (ECH) and the corresponding 
renormalisation scheme are motivated physically. 
The ECH method has been in some detail discussed in the talk of Maxwell 
\cite{maxtalk}.
For completeness here the basic formulae and results are assembled in view of 
the later interpretation of the data.
 
ECH has been originally introduced in \cite{Grunberg:1982fw}, a clear
access, taking the observable itself as perturbative expansion parameter 
(called renormalisation group improved perturbation theory, RGI)
is given in \cite{Dhar:1983py}. With respect to ECH I follow the arguments 
of \cite{Barclay:1994qa} which applies the ECH method directly to 
\epem data.
The results especially apply to mean values (or moments) of the event shape
distributions which depend on a single energy scale, the centre-of-mass energy,
$Q=\sqrt{s}$, only.

The general next-to-leading order (NLO) expression for  
event shape distributions of an observable $y$ reads:
\begin{equation}
\frac{1}{\sigma_{tot}} \frac{d\sigma(y)}{d y} = 
\frac{\alpha_s(\mu^2)}{2\pi}\cdot A(y) + 
\left (\frac{\alpha_s(\mu^2)}{2\pi}\right )^2 \cdot \left[ B(y) + 
\frac{1}{2}\beta_0
\ln(x_\mu)\cdot A(y) \right ] +{\cal{O}}(\alpha_s)^3
\label{eq:secondorder}
\end{equation}
Here $A$ and $B$ are the first and second order coefficients of the 
perturbative expansion. 
Note that due to the normalisation to the total cross-section
the integral of Eqn.~\ref{eq:secondorder} is normalised to 1.
The $\mu^2$ (or alternatively the $x_{\mu}=\mu^2/Q^2$) dependence
reflects the dependence of the prediction on the renormalisation scheme. 
In NLO
the change of renormalisation scheme is fully
equivalent to the change of the renormalisation  scale.

Weighting Eqn.~\ref{eq:secondorder} with and integrating over the 
observable then yields the mean value
$\langle y \rangle$ which can be directly measured. 
Normalising $\langle y \rangle$ to the leading order
coefficient $A$ a measurable quantity $R$ is obtained.
Such quantities are called effective charges.
The idea of the ECH method is now to chose (in any order of
perturbation theory)  a scheme 
such that all higher order coefficients for $R$ vanish.
This implies that $R$ coincides with the strong coupling extracted in a
leading order analysis and explains the name ECH as such observables 
are directly comparable to
other observables of this kind as well as to the coupling. 
In NLO this leads to the following scale:
\begin{equation}
\mu_{\text{ECH}}=Q\cdot 
e^{-\frac{1}{\beta_0}\frac{B(y)}{A(y)}}
\label{eq:xech}
\end{equation}
An effective charge $R$ obeys the Renormalisation Group Equation (RGE):
\begin{equation}
\frac{dR}{d\ln Q^2} = \beta_R(Q) \quad = \quad -\frac{\beta_0}{4}R^2\cdot
\left [ 1 + \rho_1 R + \rho_2 R^2 + \dots \right ]\quad
+ \quad K_0\cdot e^{-S/R}R^{\delta}+\dots
\label{eq:RGE}
\end{equation}
As the observables $R$ are  directly measurable in dependence of the 
energy the $\beta_R$-functions are measurable quantities. 
The leading coefficients of the $\beta$-functions, $\beta_0=11-2/3\cdot n_f$ 
and $\rho_1=\beta_1/2\beta_0$ are universal (the same as for the coupling), 
the higher coefficients are
observable dependent but free of renormalisation scheme ambiguities.
Non-perturbative contributions, invisible in perturbation theory, lead to the
term with the exponential. $K_0$ is proportional to the factor multiplying 
a power term in the usual power model expressions.
Omitting for simplicity  the non-perturbative term Eqn.~\ref{eq:RGE} can 
be integrated yielding:
\begin{equation}
\frac{\beta_0}{2} \ln \frac{Q}{\Lambda_R} =  \frac{1}{R} + 
\rho_1\ln\frac{\rho_1 R}{1+\rho_1 R} + 
\int_0^R \left \{-\frac{1}{\rho (x)}+\frac{1}{x^2(1+\rho_1 x)} \right \} dx
\quad, 
\label{eq:DG}
\end{equation}
where $\rho(R)$ is defined by the square bracketed term in Eqn.~\ref{eq:RGE}.
The constant of integration, $\Lambda_R$, should have
{\it physical significance} and 
is related to $\Lambda_{\msbar}$ by an {\bf{exact}}
expression \cite{Celmaster:1979km}. Thus \asZ\ can be calculated from
$\Lambda_R$ practically without additional uncertainty.

Besides the derivation of the above fundamental equations the paper 
\cite{Barclay:1994qa}
includes a broad discussion of the extraction of \lambdamsbar\ from \epem data
and shows the following:
\begin{itemize}
\item
The {\it ECH formalism ... is more general than the adoption of a 
particular scheme ...} It can be derived {\it non-perturbatively } assuming
only that the high energy behavior of the observable is
given by the leading order perturbative expression.
\item 
... {\it higher orders can be split into a predictable contribution}
(of known ``RG predictable'' terms $\propto \log{\mu/Q}$) 
{\it and a remaining piece containing all
the genuinely unknown aspects.}
... {\it the choice $\mu=\mu_\text{ECH}$ \it removes the predictable 
scatter and provides genuine information on the interesting ... higher-order}
terms.
\item
{\it One can exhibit the relative size of the} (truly) {\it uncalculated
  higher-order corrections for different quantities}.
These corrections {\it are related to how the energy dependence of the quantity
  differs from its asymptotic dependence as $Q^2 \to \infty$.
}
\item
{\it ... by comparing with data one can test how well the first few}
perturbative terms {\it represent the observed running.}
Marked discrepancies indicate the importance of {\it higher order terms, or
... non-perturbative contributions.}
\end{itemize}
These findings imply that the ECH scheme offers exceptional experimental
control of higher order contributions by comparing different effective charge 
observables or measuring their energy evolution. 

\subsection{Remarks on Experimental Data}
The later discussion is mainly based on precise data of shape
distributions obtained 
at the Z-peak  and of their
mean values in the full range of \epem\ experiments above the
b-threshold.
 Some more recent results 
\cite{Abdallah:2003xz,ralle,Pfeifenschneider:1999rz,MovillaFernandez:1997fr} refer to the energy range
$\sim$20-202\gev\ and comprise data on events with a 
radiated hard photon ($45 \gev < \sqrt{s´} < M_Z$) \cite{Abdallah:2003xz,ralle}.
All experimental results correspond to simple \epem\ annihilation and 
are fully corrected for experimental effects. 


It has been  realised
\cite{Salam:2001bd}, that hadron mass effects have a particularly strong
influence for the so-called jet mass observables. 
This strong mass dependence has been avoided in \cite{Abdallah:2003xz,ralle} by
redefining the particle four momenta using the so called p-scheme: 
($(\vec{p},E)\to(\vec{p},|\vec{p}|)$) or E-scheme: 
($(\vec{p},E)\to(\hat{p}E,E)$). Moreover in \cite{Abdallah:2003xz,ralle}
extra transverse momentum from 
B-hadron decays was corrected for by using Monte Carlo.
This correction is power behaved and can have influence on the energy
evolution of the observables. A figure displaying the correction is included
in \cite{Abdallah:2003xz}. 

B-decays also influence event shape
distributions. For 2-jet like topologies the distributions are depleted. 
These events are shifted towards 3-jet topologies by the extra transverse
momentum from the decay. For observables calculated from the whole
event (like $B_{T}$) the decay from the narrow event side contributes
for all values of the observable leading to visible effects in the whole range
of the observable. In principle, the differences between the observables
due to B-decays need to be
corrected when applying power correction models to shape distributions.
Plots of such corrections are again included in \cite{Abdallah:2003xz,ralle}.

\section{Results on Event Shape Distributions}
\subsection{Effects of Changing the Renormalisation Scale}
\newlength{\wi}   \wi 0.48\textwidth
\newlength{\fwi}  \fwi 0.9\wi
\newlength{\hwi}  \fwi 0.9\wi
\begin{figure}[hb]
\begin{minipage}[b]{\wi}
\centering\includegraphics[width=\fwi]{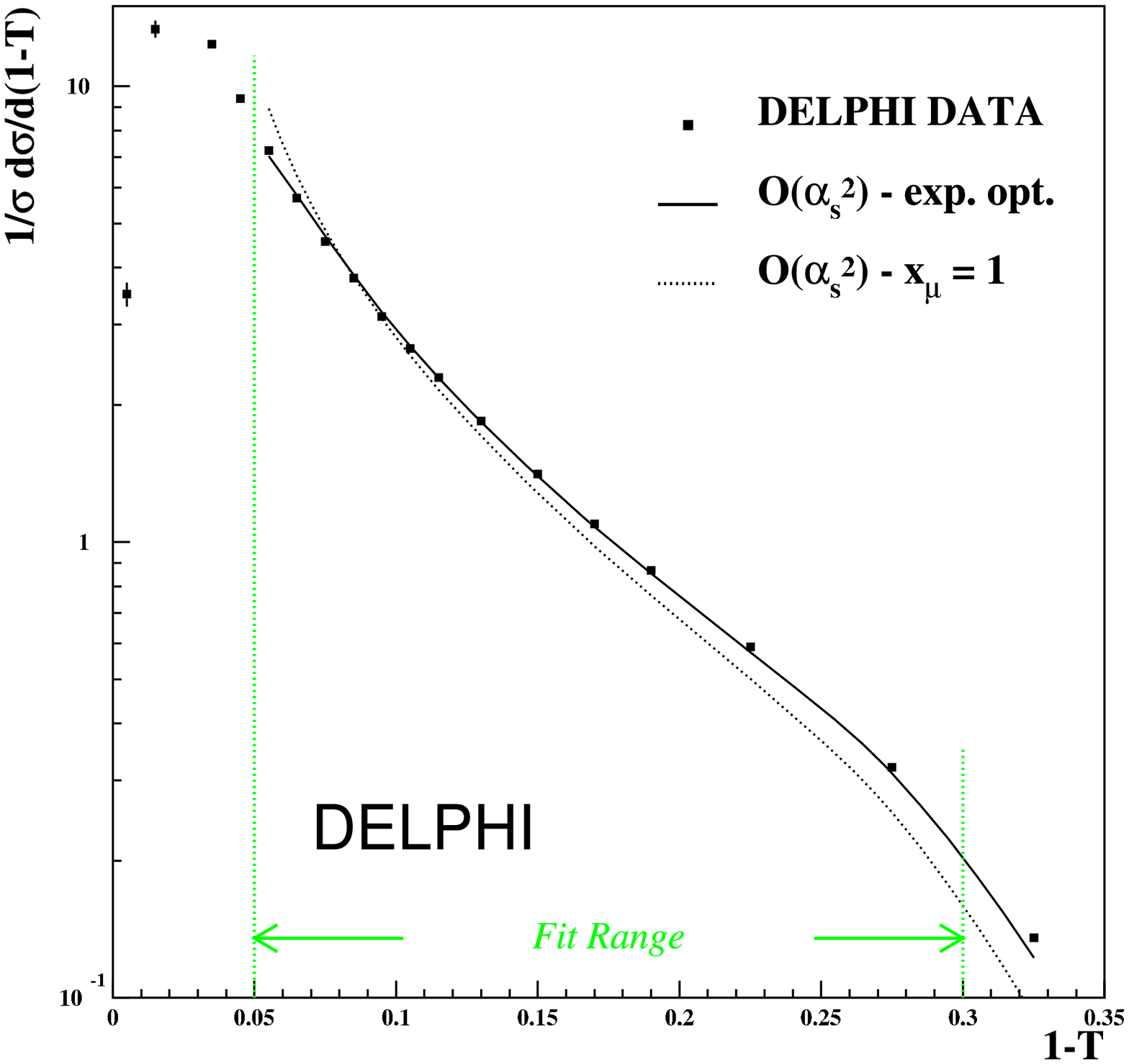}
\caption{\oassq\ prediction for $\tau=1-\text{Thrust}$ for \xmu=1 
and for the experimentally optimised value of \xmu.}
\label{f:thrxmu}
\end{minipage}\hfill
\fwi 0.72\wi
\begin{minipage}[b]{\wi}
\centering\includegraphics[width=\fwi]{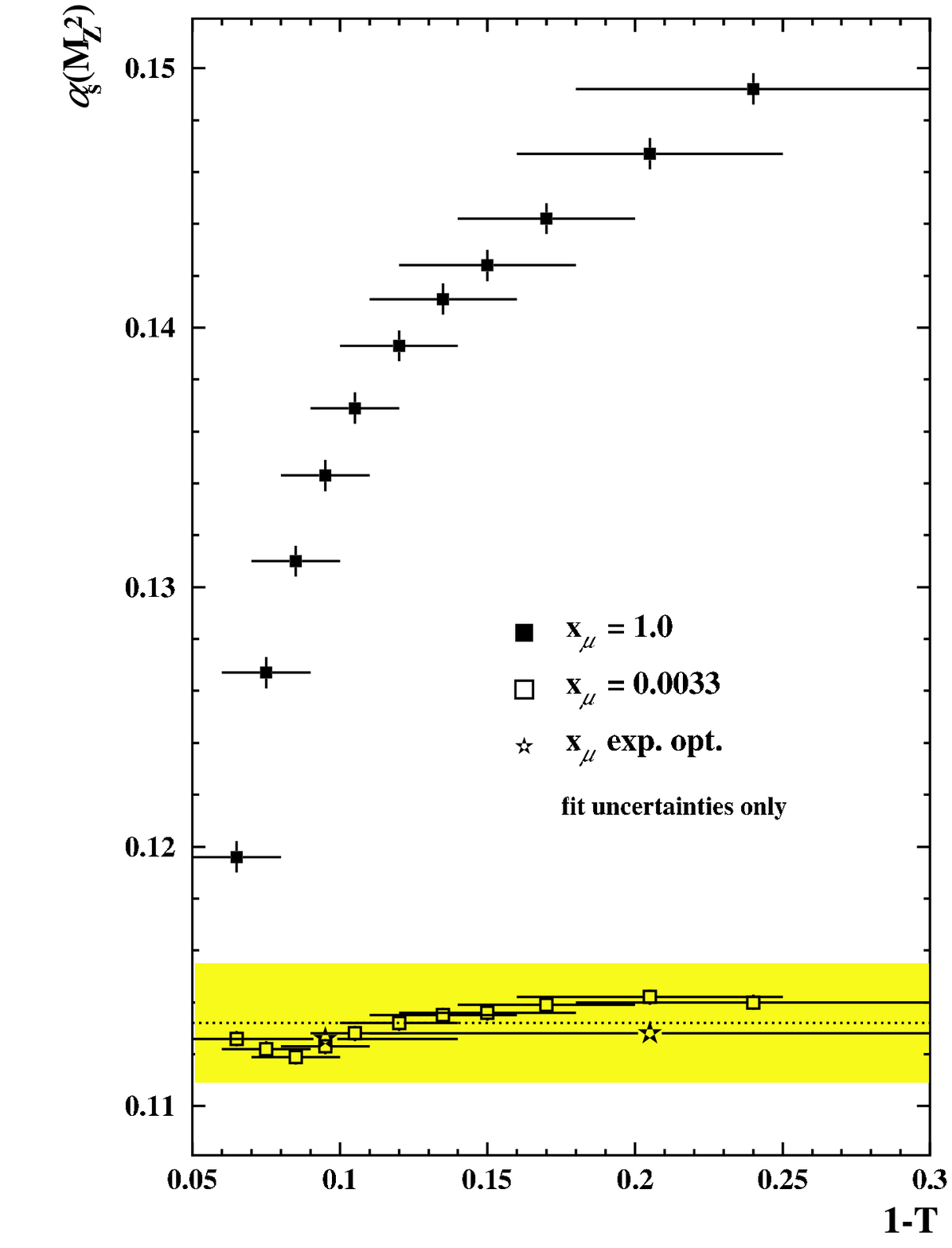}
\caption{\asZ\ from $\tau=1-\text{Thrust}$ as function of $\tau$ for \xmu=1 
and for the experimentally optimised value of \xmu.}
\label{f:asthrxmu}
\end{minipage}
\end{figure}
Analyses of event shape distributions performed by experimental collaborations 
mainly employed
the \msbar-scheme with $x_{\mu}=1$; information about the explicit use of the 
ECH scheme \cite{Abreu:2000ck,siggi,Burrows:1995vt} is sparse. 
Several experiments attempted,
however, to determine optimal scales by fitting \as\ and \xmu\ simultaneously
to the data. In view of this, so called experimental optimisation, it is
instructive to discuss the influence of \xmu\ on the prediction
Eqn.~\ref{eq:secondorder}. 
In Fig.~\ref{f:thrxmu} the data on $\tau=1-\text{Thrust}$ is compared with
the prediction Eqn.~\ref{eq:secondorder} for $\xmu=1$ and the optimised scale
($\xmu\sim 0.0033$). Evidently the change of scale leads to a 
``turn'' of the prediction. For $\xmu=1$ the slope of the data is barely
described. Other observables show a similar though often less pronounced
discrepancy. 
As a consequence the quality of the fits for \xmu=1
is, in general, unsatisfactorily bad and the fitted values of \as\ depend
on the
interval chosen for the fit. This is shown in Fig.~\ref{f:asthrxmu} e.g. for
the Thrust \cite{siggi}. 

This behavior is basically known since \petra\ \cite{Magnussen:1989eh} or
the early days of \lep\ and has been observed at small values
of the event shape observables (in the 2-jet region)  
where the perturbative \oassq\ predictions 
are presumed little reliable. It should be noted, however,  that due to the
normalisation of the event shape distributions $1/\sigma \int d\sigma/dy~dy=1$,
the misfit between data and prediction must persist at large $y$ if present
at small $y$. This has indeed been seen \cite{Abreu:2000ck,siggi}. 

The optimisation of a single scale value for a distribution has been
criticised \cite{Barclay:1994qa} as, in principle, the scale is expected 
to be $y$-dependent (compare Eqn.~\ref{eq:xech}).
However, the change of scale expected in the typical fit ranges of the data
is moderate (compare Fig.~\ref{f:correl}). 
Experimental optimisation therefore presents a fair 
compromise between 
theoretical prejudice or request and experimental feasibility.

Results on experimental optimisation are included in \cite{Abreu:2000ck,Burrows:1995vt,Abreu:1992yc,Acton:1992fa}.
The first reference, which is discussed below, comprises 18 observables 
and is based on high statistics
data. Moreover in this analysis the dependence of the event shape
distributions on the polar angle of the event axis with respect to the beam 
has been exploited. This is advantageous in view of experimental systematics
and leads to a large number of statistically independent data points.

\pagebreak
\noindent
{\bf Consequences for Matched NLLA {\boldmath\oassq}\ Analyses}\\
\indent 
The mismatch of the slope of data and \oassq\ prediction has consequences
also for the analysis of event shape distributions with matched NLLA/\oassq\
predictions.
\fwi 0.9 \textwidth
\begin{figure}[tb]
\centering\includegraphics[width=\fwi]{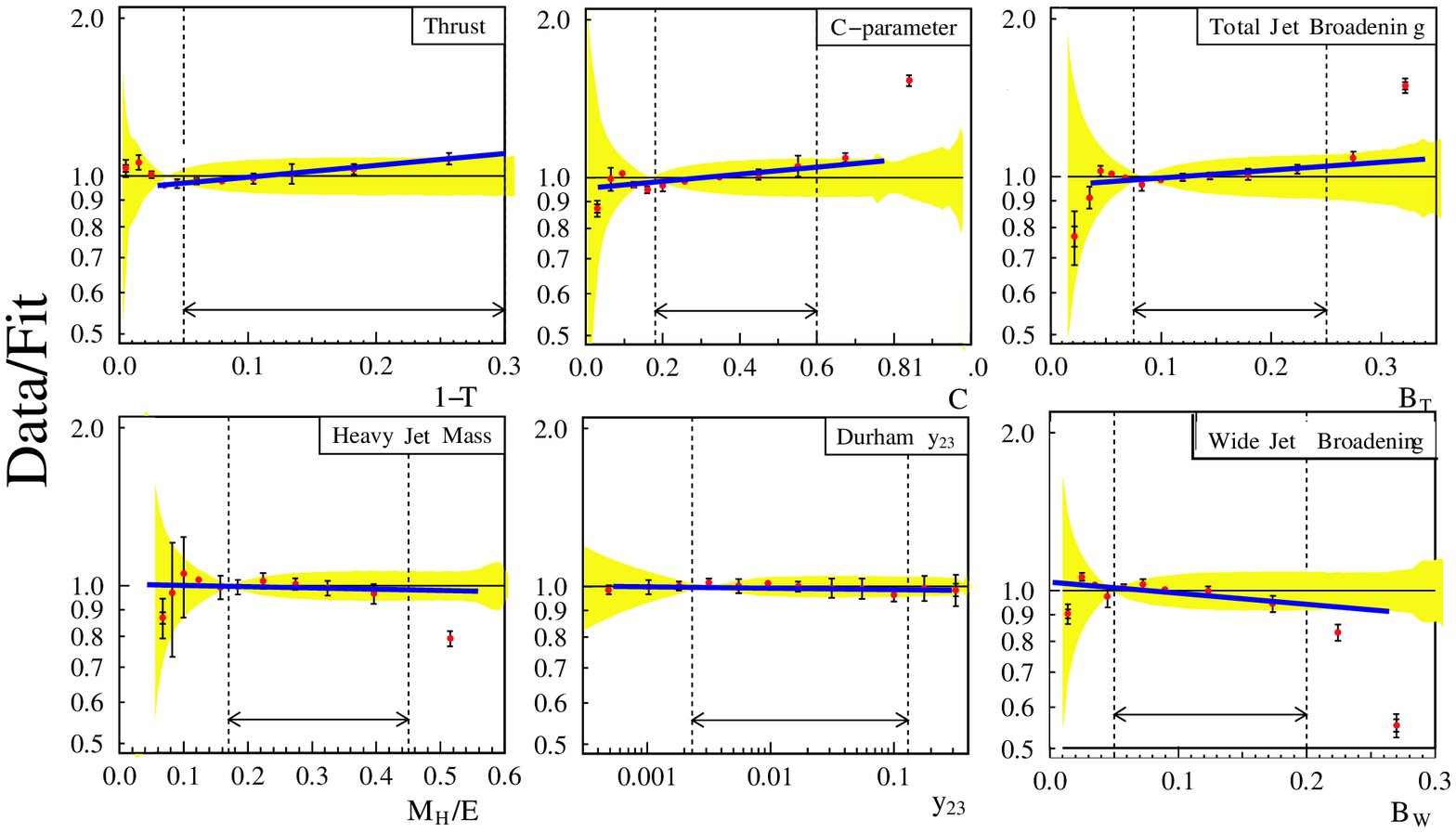}
\caption{Ratio data over NLLA/\oassq-fit from \opal\ 
\cite{Abbiendi:2004qz,Ford:2004dp}
(lines added by the author).}
\label{f:opalratio}
\end{figure}
The matching 
from an
experimental point of view represents 
a kind of ``averaging'' of both
predictions where in the 2-jet regime the NLLA part and in the 3-jet regime
the \oassq\ part dominates. As in the matching the  \oassq\ prediction is used
with a renormalisation scale value \xmu=1 the slope of the data is,
for many observables, imperfectly described. In consequences the same problems
as discussed above for the pure \oassq\ prediction persist, though diminished
by the matching with the NLLA part. In consequence the $\chi^2/N_{df}$ is 
often in-acceptably bad when applying standard rules. 
Additional theory errors are introduced to regain consistency.
As in the \oassq\
case the fitted value of \asZ\ often shows a marked dependence on the
observable. To illustrate this in Fig.~\ref{f:opalratio} the ratio of the
data and the fitted matched predictions is shown for several observables 
from the
concluding \opal\ analysis on event shape distributions
\cite{Abbiendi:2004qz,Ford:2004dp}.
In order to make the discrepancies directly
visible straight lines are additionally put to the data points.
For $T$, $C$ and $B_{T}$ a clear increase of the ratio with increasing values
of the observables is seen corresponding to an increase of \as\ ($\gtrsim
5\%$).
For $M_H/E$ and $y_{23}$ the ratio is almost
constant, while for $B_{W}$ the ratio decreases with increasing observable.
This more complicated pattern 
correlates with the importance of higher order corrections in the \oassq\ part
of the prediction or simply with the ratio $B/A$.
The three observables with increasing ratio also show a large ratio $B/A>15$,
for $M_H/E$ and $y_{23}$ the ratio is smaller. For $\langle
B_{W} \rangle $  it is negative. The mismatch of the slope of the 
data and the prediction also deteriorates the initial $\chi^2/N_{df}$.
The constructed error bands (dominated by ``theory errors'') restore 
consistency of the fits.
The best stability is obtained for
$y_{23}$ and justifies the small error obtained for this observable
(see Fig.~\ref{f:opalratio} and 
\cite{Abbiendi:2004qz,Ford:2004dp,Heister:2002tq}). 

Still the \asZ\ values depend on the fit range for some 
observables and are thus biased.
This observable dependent biases is even present in the
overall \lep\ combination 
(see Fig.~\ref{f:aslep}, \cite{QCDwg}). 
The (correlated) \as\ values from $T$ and $B_T$ are high, 
$y_{23}$ and $B_{W}$ are low 
compared to the average. $C$ is close to the average and 
seems to form an exception.

It is likely that the imperfect description of the slope by the matched 
prescription similarly 
influences analyses of power terms introduced in order to describe the 
hadronisation. 
E.g. in the corresponding \delphi\ analysis \cite{Abdallah:2003xz}
the \as\ values for $B_T$ and $\rho_h=M^2_h/s$ (E-scheme) are markedly smaller
compared to the results from $B_{W}$, $T$ and $\rho_{s}$.

\subsection{Results from Distributions on Experimentally Optimised Scales}
\wi 0.48\textwidth
\fwi 0.8\wi
\hwi 0.88\wi
\begin{figure}[hb]
\begin{minipage}[b]{\wi}
\centering\includegraphics[width=\hwi]{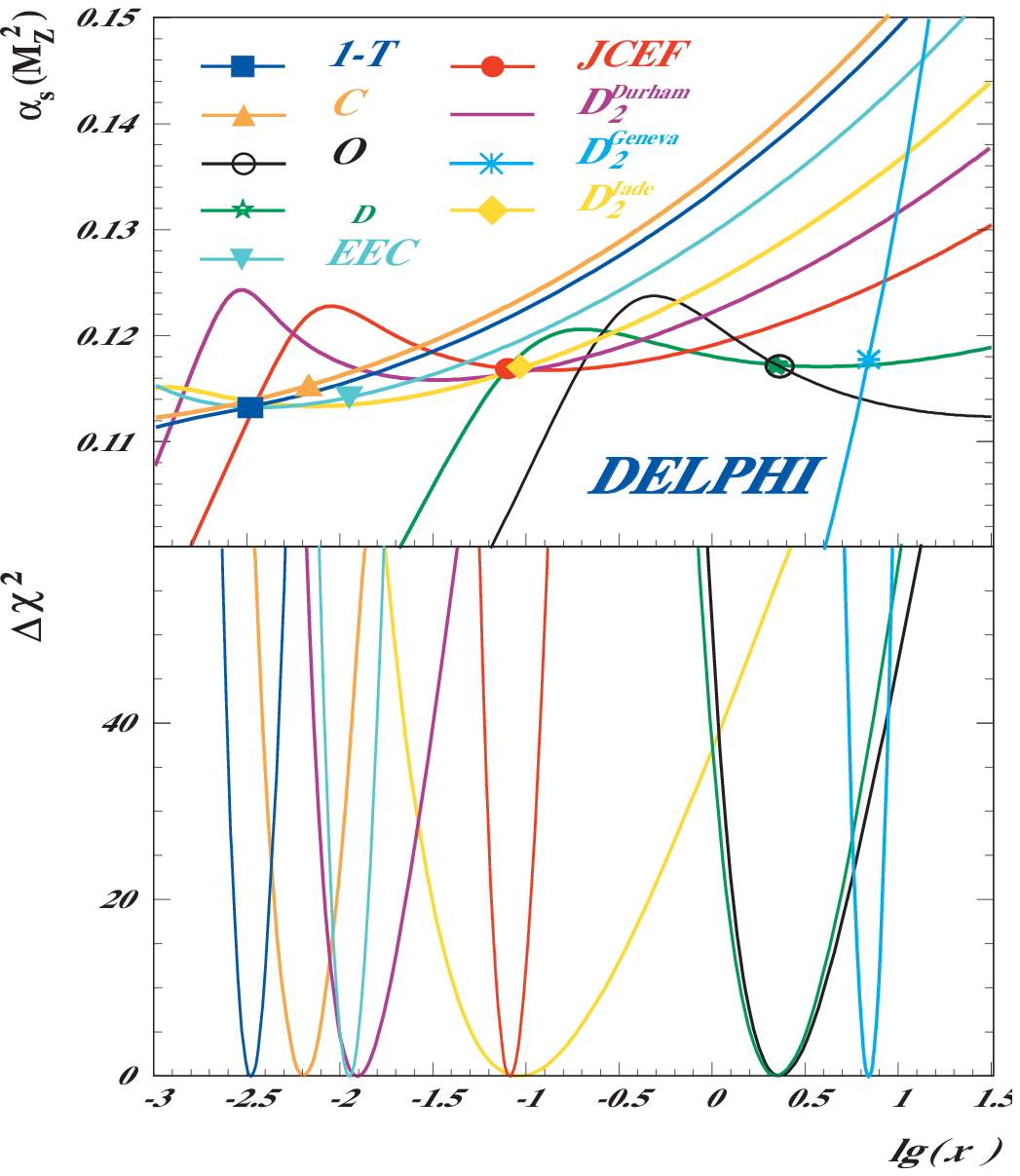}
\caption{$\Delta\chi^2$ and \asZ\ obtained from \oassq\ fits of 18 event shape
distributions as function of \xmu\ \cite{siggi}.}
\label{f:chiasxmu}
\end{minipage}\hfill
\fwi 0.8\wi
\begin{minipage}[b]{\wi}
\centering\includegraphics[width=\fwi]{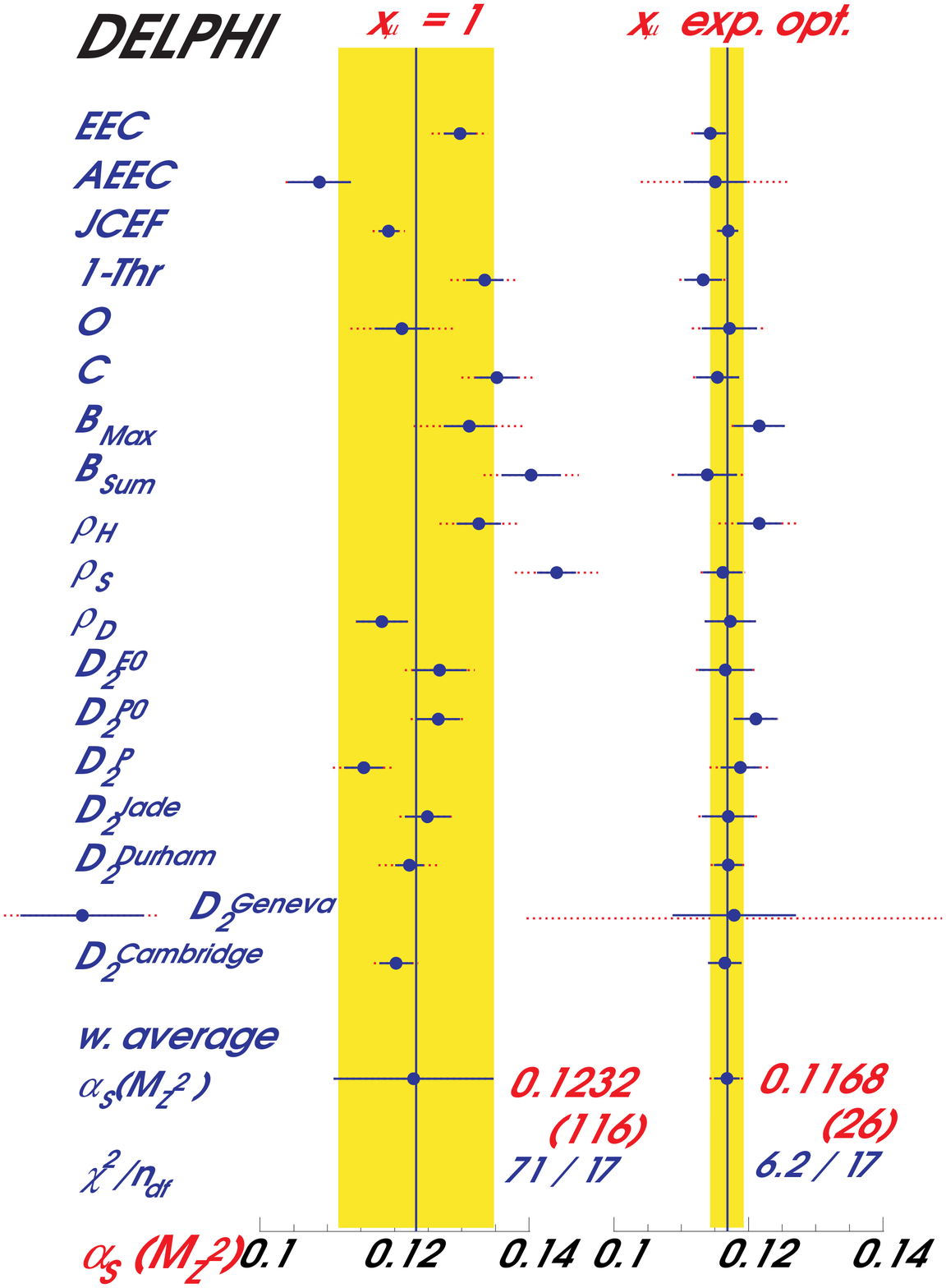}
\caption{\asZ\ results obtained from \oassq\ fits of 18 event shape
distributions for \xmu=1 and \xmu=\xmuopt.}
\label{f:asboth}
\end{minipage}
\end{figure}
In Fig.~\ref{f:chiasxmu} the $\Delta \chi^2$ and \asZ\ as function of \xmu\ as
obtained from several event shape distributions is shown
\cite{Abreu:2000ck,siggi}. The fit quality for these fits is satisfactory when
including experimental and hadronisation errors only. Marked minima in 
$\Delta \chi^2$ are observed scattered over a wide range of \xmu.
The clear minima imply that the optimised scale values
are statistically well determined. The \asZ\ results corresponding to these
scales show a much smaller scatter as for fixed \xmu=1. This is more clearly
shown in Fig.~\ref{f:asboth}. The spread reduces from $9.4\%$ for \xmu=1
to 2.2\% with optimised scales, \xmuopt. 

\fwi .6\textwidth
\begin{figure}[ht]
\centering\includegraphics[width=\fwi]{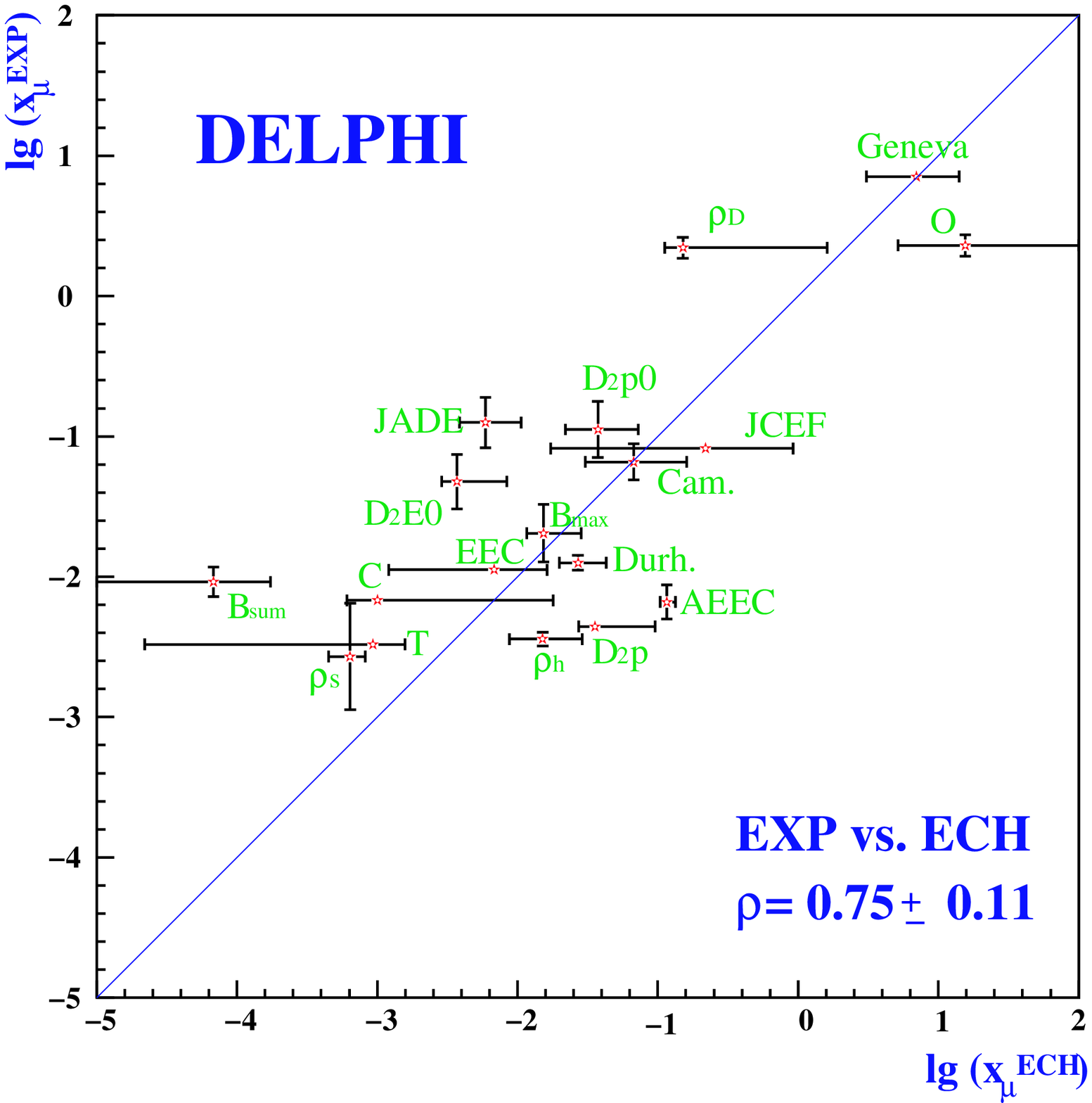}
\caption{Correlation of the experimentally optimised scale \xmuopt\ values with
the ECH expectation $x_{\mu}^{ECH}$. The errors of \xmuopt\ are fit errors,
for $x_{\mu}^{ECH}$ they indicate the expected change in the fit range.}
\label{f:correl}
\end{figure}
It is particularly interesting to study the
correlation of the fitted \xmuopt\ with the values 
expected from ECH, \xmuech, shown in
Fig.~\ref{f:correl}. The \xmuopt-errors are from the fit, the error bars
 for \xmuech\ indicate the range of scales expected within the fit
interval of the data. A significant correlation $\rho=0.75\pm 0.11$ is
observed.
The wide spread of experimentally optimised scales which has been often criticised is in
fact slightly smaller than the range expected from ECH. 

Regarding the distribution of
individual observables (indicated by the letters in the plot) it is seen that
observables calculated from the whole event (e.g. the Thrust) tend to populate
the region of small \xmuech, observables sensitive mainly to the wide side of
the events (e.g. $\rho_h$) populate the center while observables from
the difference of the wide and narrow side show large scales. This ordering 
corresponds (via Eqn.~\ref{eq:xech}) to large (for small \xmuech) and small 
 second order \msbar\ contributions (for big \xmuech).
Overall the observed large range of \xmuopt\ can be considered as 
understood in ECH theory.

It may be worth noting that the RMS spread%
\footnote{Estimated from the data of Fig.~\ref{f:correl} as
  $10^{\sigma(P(\lg(x_{\mu}^{ECH}/x_{\mu}^{opt})))}$. } 
of \xmuopt\ with respect to
\xmuech\ is only 2.5 (i.e. similar to the 
range used for naive \msbar\ theory error estimates). This may be taken as a
justification of an ECH theory error estimate by a corresponding scale change. 
However, the corresponding \as\ uncertainty will be far smaller
than in the \msbar\ case as
the optimal scales often correspond to the minima of the \asZ(\xmu) curves
(compare Fig.~\ref{f:chiasxmu}). 

\subsection{Results from Event Shape Mean Values}
\wi \textwidth
\fwi 0.73\wi
\begin{figure}[p]
\centering\includegraphics[width=\fwi]{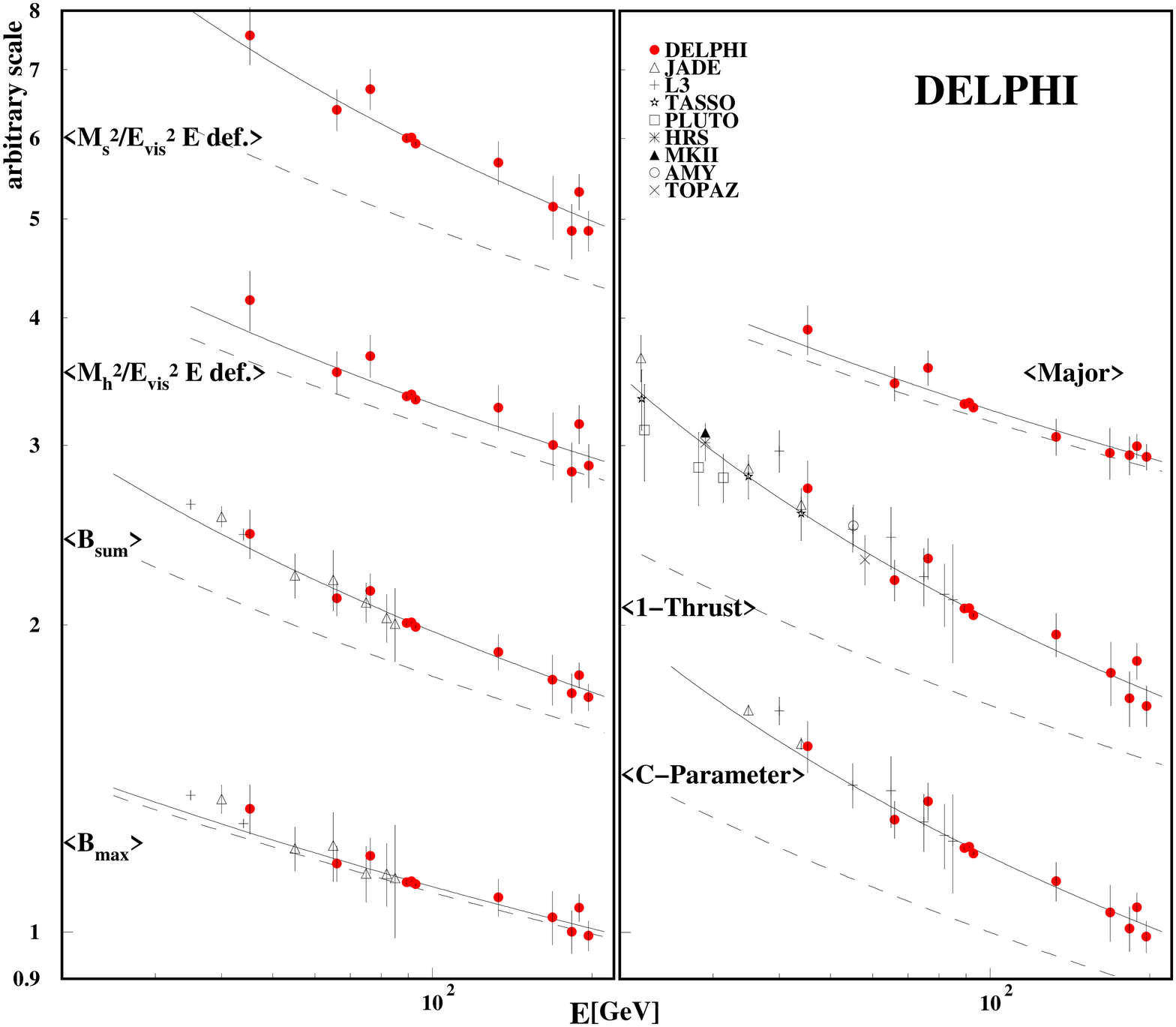}
\caption{RGI fits to several event shape means as function of the
  centre-of-mass energy (full line). The dashed line represents the
  \msbar-expectation (with the same \asZ). }
\label{f:rgifits}
\vfill
\centering\includegraphics[width=\fwi]{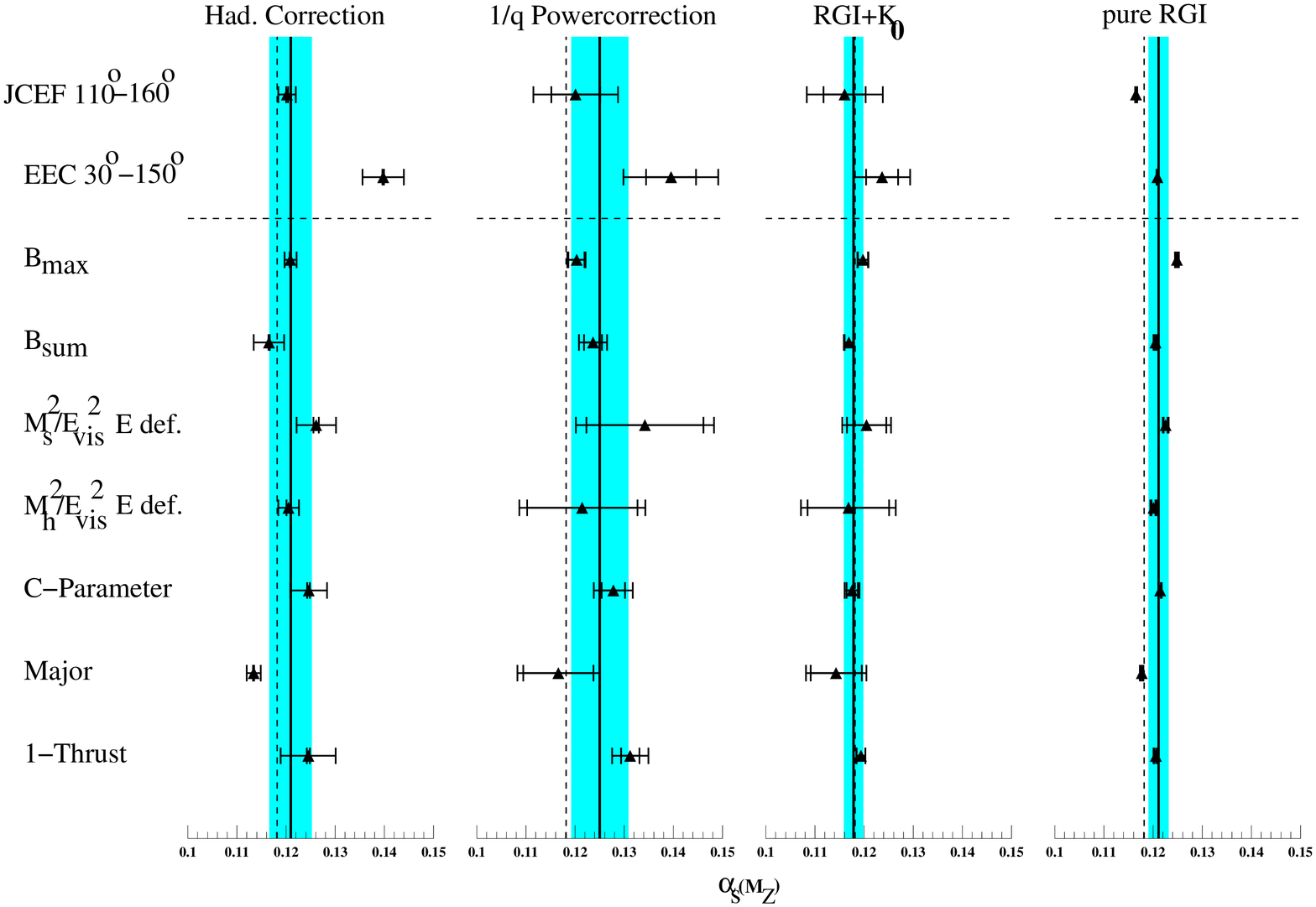}
\caption{\asZ\ results obtained from several event shape means or integrals
over the JCEF or EEC using \msbar\ theory plus Monte Carlo  or power 
hadronisation corrections and RGI theory with and without
non-perturbative terms.}
\label{f:rgias}
\end{figure}
Some relevant data on mean values of shape distributions as function of the
centre-of-mass energy \cite{Abdallah:2003xz} is shown in
Fig.~\ref{f:rgifits}. 
The data is compared
to fits of Eqn.~\ref{eq:DG} shown as full lines. The dashed lines represent
the \msbar-prediction for the same \asZ.
It turned out that the data can be well described by 
the ECH/RGI fits  with a
single value of \asZ\ ($\sim 0.120$) and negligible additional non-perturbative terms.
In Fig.~\ref{f:rgias} the \asZ\ results obtained from this data with
\msbar\ prediction and Monte Carlo  or power hadronisation corrections
and ECH/RGI theory with and without power terms are compared.
Evidently the spread of the results is far smaller for the ECH/RGI results.
The inclusion of power terms leaves the spread almost unchanged
but lowers the average value of \asZ\ to $0.118$. 
This result implies that the differences due to non-trivial uncalculated higher
order terms or non-perturbative terms is small, of ${\cal{O}}(2\%)$. 
Note that higher order corrections differing by about a factor 2 
(in \msbar) are 
expected~\footnote{Compare e.g. the $B/A$ ratios in Fig.~14 of 
\cite{Abdallah:2003xz}. }
 for event shape observables calculated from the full event or the wide 
hemisphere of an event only. This is also reflected e.g. in the perturbatively
calculated ratio of the power terms of Thrust and $\rho_h$.
The bigger differences seen
in the \msbar\ compared to the ECH results then should be due to 
``RG predictable'' $\log \mu/Q$ terms.

\begin{figure}[tbh]
\wi 0.16\textwidth
\hwi 0.4\textwidth
\fwi 0.58\textwidth
\begin{minipage}{\fwi}
{\centering\includegraphics[width=\fwi]{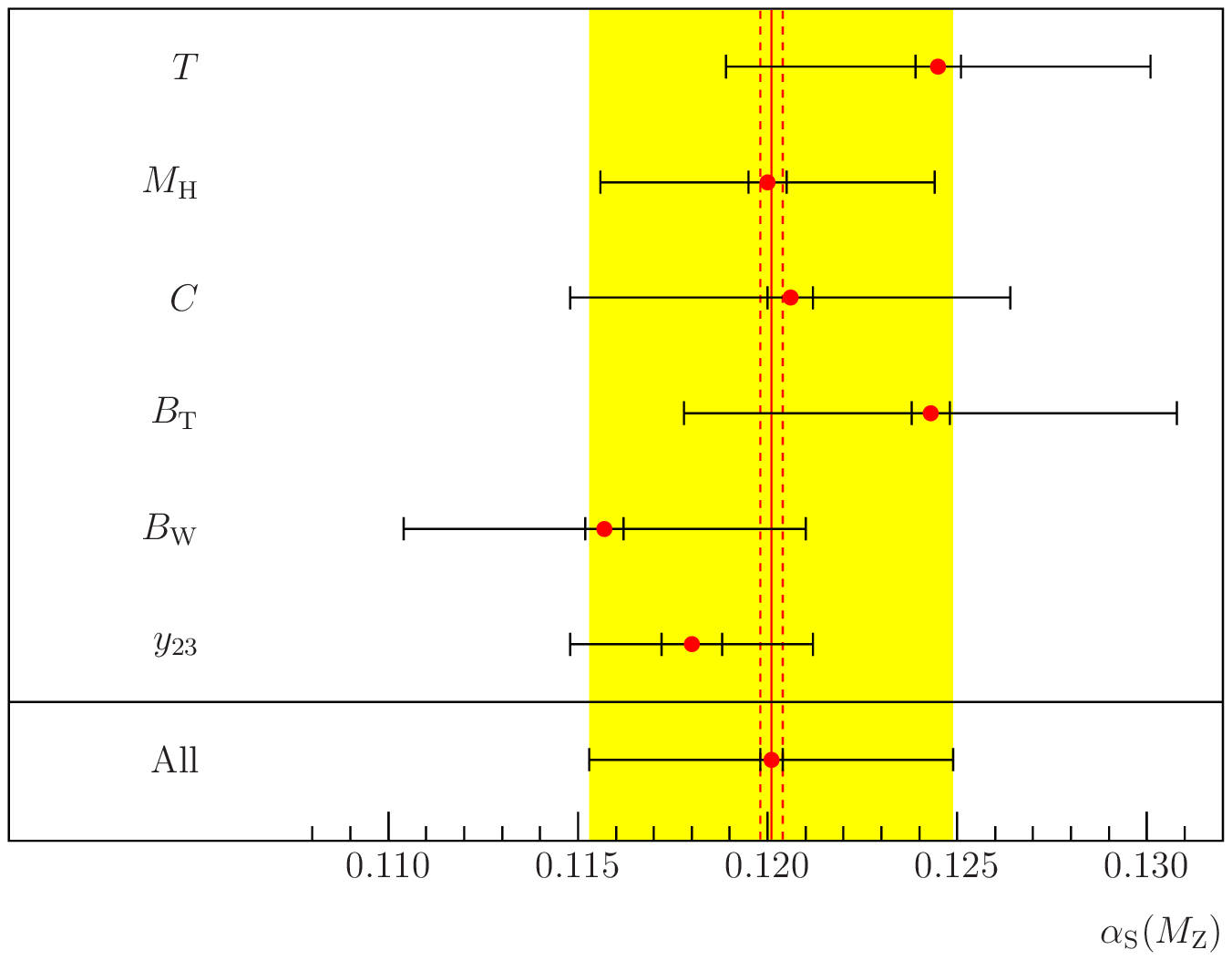}}
\caption{Combined \lep\ \asZ\ results for different observables 
\cite{Ford:2004dp}. \label{f:aslep} }
\end{minipage}\hfill
\begin{minipage}{\hwi}
{\centering\includegraphics[width=\wi]{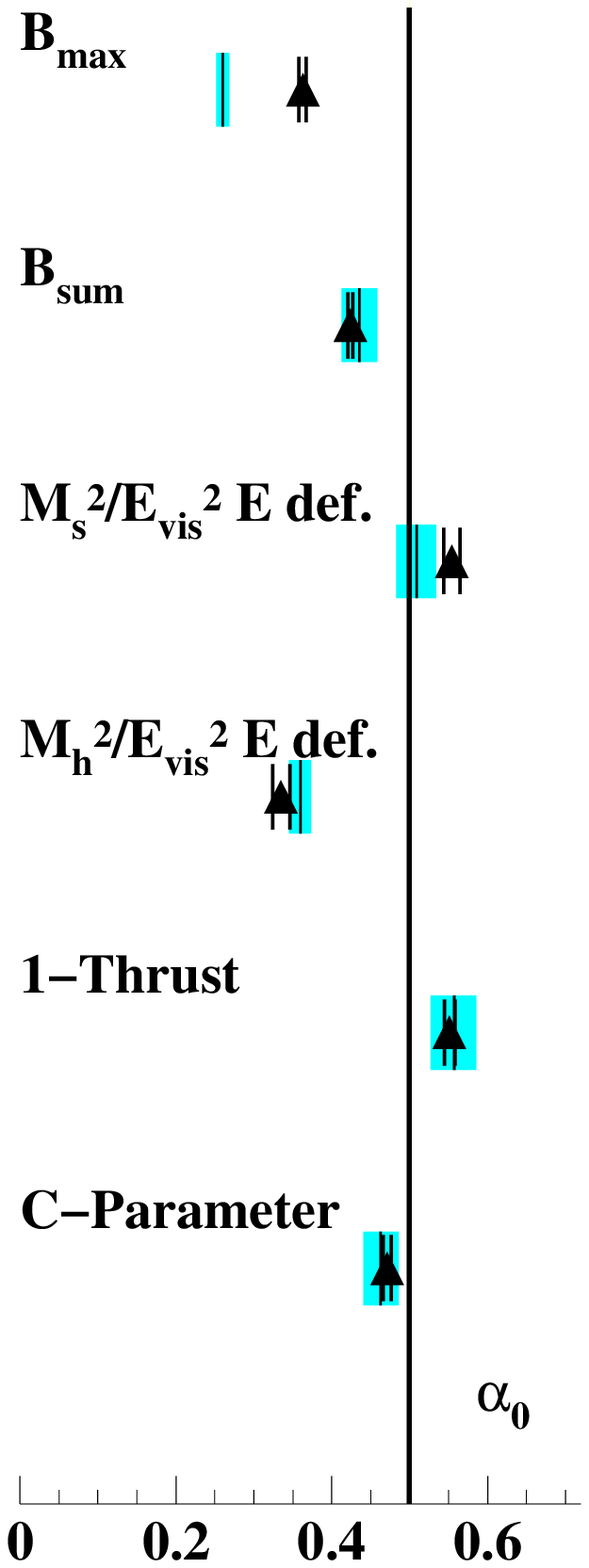}}
\caption{Size of the non-perturbative terms $\alpha_0$  as determined from
  power model fits compared to ECH expectation (grey bands). }
\label{f:a0rgi}
\end{minipage}
\end{figure}

The better agreement of the ECH result is also illustrated by
Fig.~\ref{f:a0rgi} where the measured size of the
power terms for some observables as parameterised by $\alpha_0$  is 
compared to the ECH/RGI expectation which has been calculated by
setting the \msbar+power terms expression equal to the ECH prediction.
It is evident that the RGI/ECH prediction describes the data better than 
a universal value of $\alpha_0$ presumed by power correction models.

\begin{table}[b]
\caption{Results on $\beta_0$ obtained from RGI/ECH fits to 
\delphi\ data \cite{Abdallah:2003xz}.
The first error is statistical, the second is systematic and the 
third is due to the B-hadron decay correction. 
The uncertainty due to possible non-pert. contributions is small.}
\label{t:bet0cmp}~\\
\begin{center}
\begin{tabular}{|l|c|c|}
\hline \textbf{Observable} & $\mathbf{\beta_0}$ & $\mathbf{\chi^2/N_{df}}$ \\
\hline 
$\langle 1-\text{Thrust} \rangle$     & $7.7\pm1.1\pm0.2\pm0.1$ & $9.2/13$\\
$\langle \text{C-parameter} \rangle$  & $7.8\pm1.0\pm0.3\pm0.1$ & $7.2/13$\\
$\langle \rho_h \rangle$ E-def.& $7.5\pm1.5\pm0.2\pm0.0$ & $8.8/13$\\
$\langle \rho_s \rangle$ E-def.& $7.5\pm1.1\pm0.2\pm0.0$ & $7.1/13$\\
$\langle B_{\text{W}}\rangle$         & $7.7\pm1.4\pm0.1\pm0.1$ & $6.3/13$\\
$\langle B_{\text{T}}\rangle$         & $7.7\pm0.9\pm0.1\pm0.1$ & $5.9/13$\\
$\langle {\text{Major}}\rangle$       & $8.0\pm1.1\pm0.1\pm0.1$ & $9.2/13$\\
\hline
\end{tabular}
\end{center}
\end{table}

Besides the comparison of the \as\ results from different observables
the energy dependence of these observables i.e. the observed
$\beta_R$-functions
can be compared to the NLO expectation.
As the NLO terms present only a small
${\cal{O}}(4\%)$ correction  $\rho_1$ has been set to the QCD expectation.
Then the $\beta_0$ values obtained from the different observables 
can be directly
compared. In order to allow full control of the systematic uncertainties this
comparison was made for seven observables determined from 
\delphi\ data only. The measured $\beta_0$ values
agree among each other within part of the correlated 
statistical uncertainty and with the QCD expectation $\beta_0=7.66$
(see Tab.~1). The possible influence of power terms ($\sim 2\%$ at
the Z) on the $\beta$-function is small as the energy dependencies of the 
power terms and the $\beta$-function are similar.

\begin{figure}[tb]
\begin{minipage}{\hwi}
\centering\includegraphics[width=\fwi,height=9.5cm]{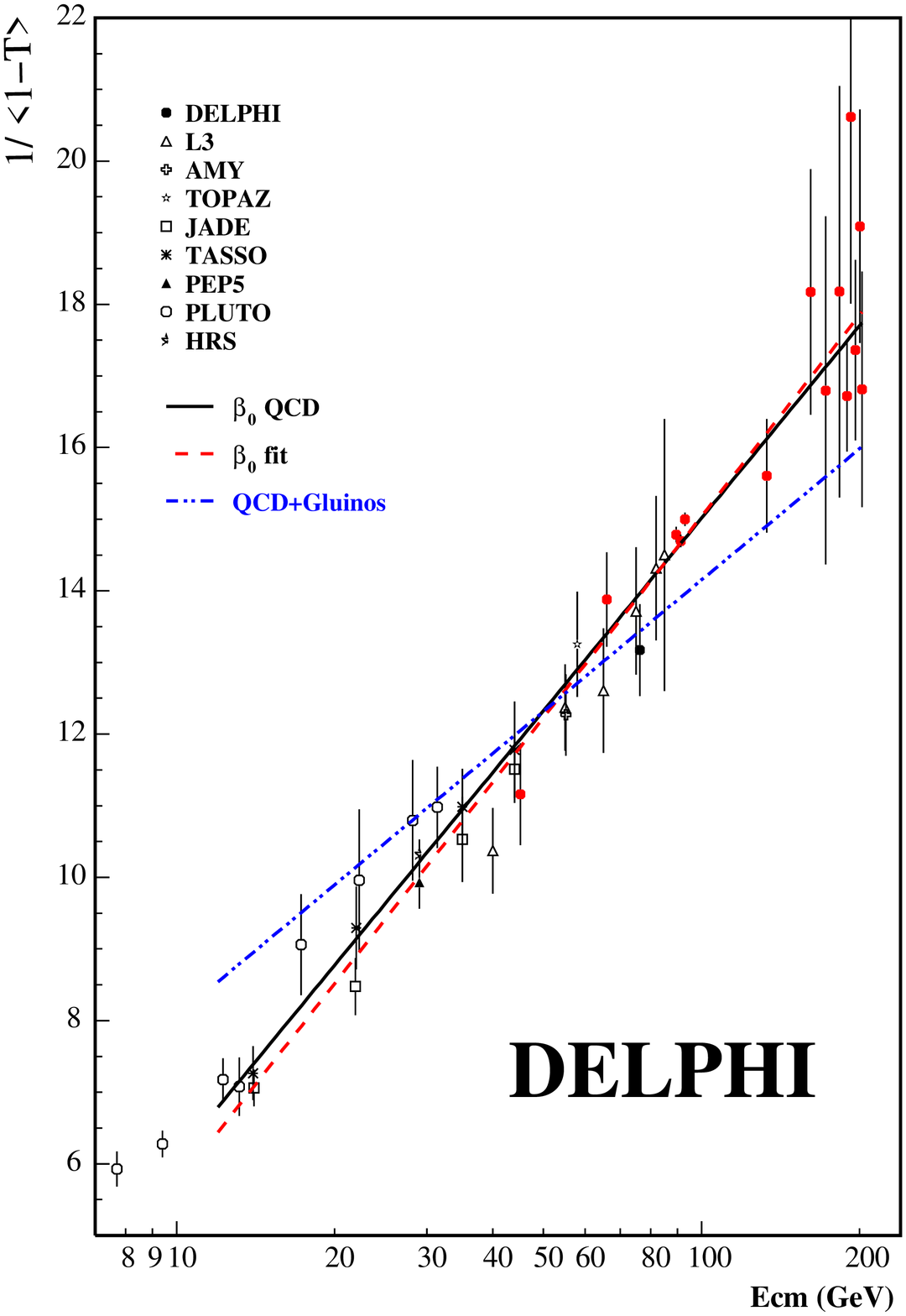}
\caption{ECH/RGI fit to data on $\langle 1-\text{Thrust} \rangle$.
The data is corrected for the small influence of B-hadron decays.
The full line represents the QCD expectation, the dashed-dotted line the
expectation for QCD plus light Gluinos.}
\label{f:rgithrustfit}
\end{minipage}\hfill
\begin{minipage}{\hwi}
\centering\includegraphics[width=\fwi]{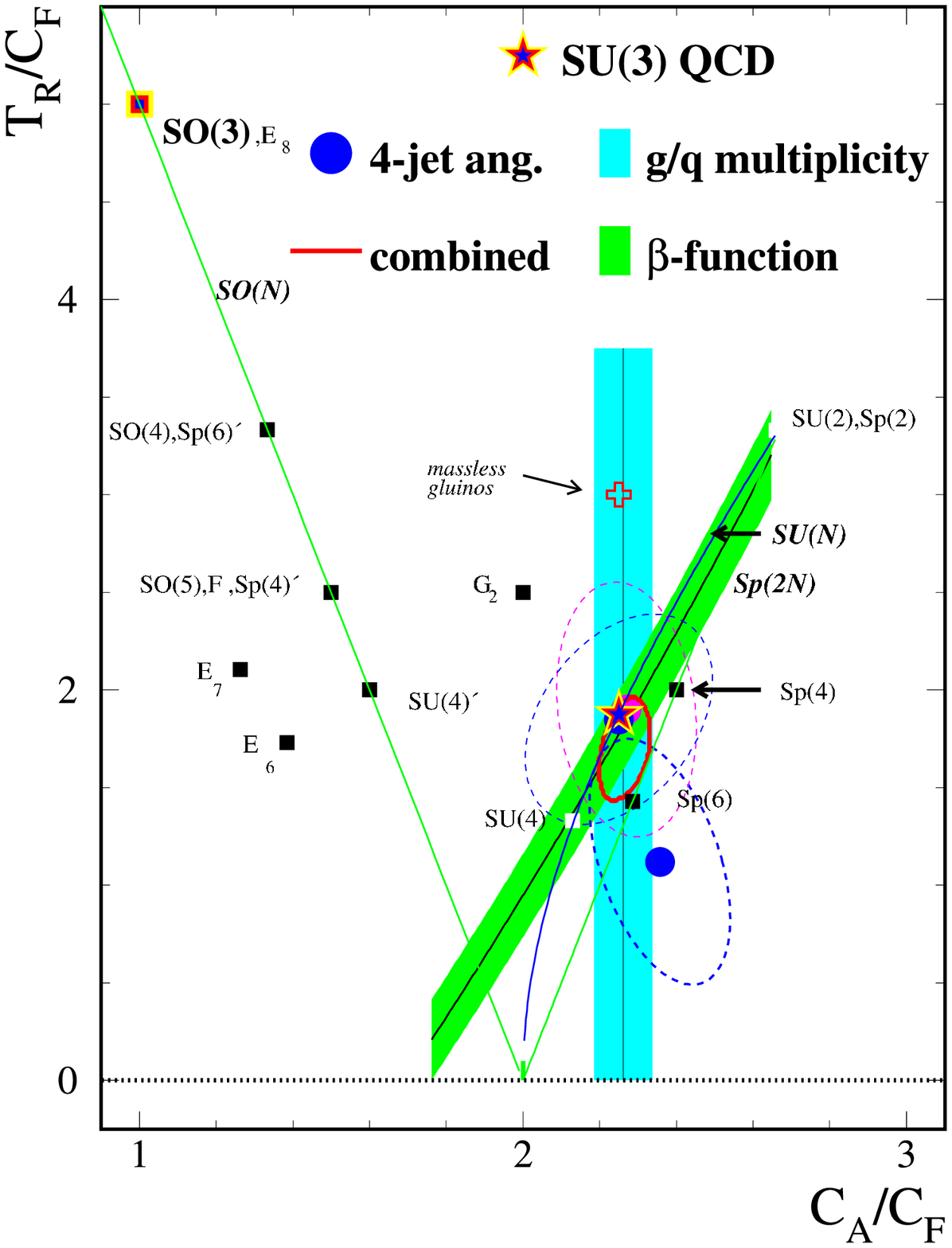}
\caption{Constraints on $C_A/C_F$ and $T_R/C_F$ from measurements of the
  $\beta$-function, the multiplicity ratio in gluon to quark jets and four jet
  angular distributions.}
\label{f:wimplot}
\end{minipage}
\end{figure}

Especially in view of a measurement of the $\beta$-function 
the analysis has been
repeated including reliable low energy data on the Thrust. 
The resulting fit is shown
in Fig.~\ref{f:rgithrustfit}, and corresponds to 
\begin{xalignat}{1}
\beta_0 & = 7.86  \pm 0.32 \quad.
\end{xalignat}
This measurement of the $\beta$-function is the most precise presented so far
and allows to strongly constrain the QCD gauge group to SU(3)  
in combination with measurements of the multiplicity ratio 
of gluon and quark jets \cite{Abdallah:2005cy} and four jet angular 
distributions \cite{Heister:2002tq, Abreu:1997mn, Abbiendi:2001qn} 
(see Fig.~\ref{f:wimplot}).
This agreement in turn lends further support to the measurements using the ECH scheme. 
\wi \textwidth
\fwi 0.43\wi
\hwi 0.47\wi

As a final check it has been tried to infer the possible size of $\rho_2$
from the overall Thrust data by fitting the corresponding expression 
with $\beta_0$
and $\rho_1$ fixed to the QCD expectation. The resulting value of
$\rho_2$ was found to be small and consistent with 0.

The above quoted results imply that the running of \epem data already
in the energy range above the b-threshold is well described by the leading
two coefficients of the QCD $\beta$-function expansion. Neither important
NNLO contributions nor important non-perturbative terms are observed from
the running of the observables or from the comparison of \as\ obtained from
different observables.
In turn this gives confidence to the ECH/RGI measurements of 
\asZ\ from event shape 
means.

\section{Summary}
The extraction of \asZ\ from seven event shape distribution mean values 
(corrected for mass effects) using
the ECH/RGI formalism shows a far better consistency between the individual
results (spread $\sim 2\%$) compared to the ones  obtained using the
conventional \msbar\ \oassq\ analysis combined with Monte Carlo or power model
hadronisation corrections (spread $\sim 7\%$). Moreover the energy evolution
of these observables, the $\beta_R$-functions are well described
by the two-loop QCD expression and a universal value of \asZ.
Possible non-perturbative contributions turn out to be small ($\sim 2\%$ at the
Z). A fit to the precise data on Thrust in the energy range 15-205\gev\ 
precisely confirms the two loop QCD  expectation.

These experimental result support the theoretical prejudice that the ECH
scheme
is more general than a particular renormalisation scheme as it can be derived
non-perturbatively. Moreover large logarithmic terms $\propto \log\mu/Q$ are
avoided in ECH. This fact and the possibility to directly measure the
$\beta_R$-functions and thereby to judge the importance of genuine 
higher order and
non-perturbative terms single out the ECH scheme for experimental tests.

The inciting results obtained for mean values are corroborated by 
studies of event shape distributions with experimentally optimised scales.
These scales turn out to be similar to the ECH expectation. 
Also here the spread of 
the \asZ\ results is strongly reduced although additional logarithmic
dependencies  on the observables are to be expected.

In studies employing \msbar\ theory and the so-called physical scale \xmu=1
the slope of the distribution is often badly described. This leads to
a dependence of the extracted \asZ\ value on the fit range, thus to a biased
result.
Due to the normalisation of the event shape distributions to one 
such a dependence 
must be present, 
though to reduced extent, in the state-of-the-art matched 
\oassq/NLLA analyses and most likely causes part of the spread
of the \asZ\ results extracted using this prescription. 
Despite the technical
problems encountered \cite{maxtalk}  studies to match ECH and NLLA
calculations should therefore proceed.

In the concluding talk of the workshop  \cite{muller} the question was raised
why both ECH and \msbar/power models lead to a reasonable description of the
energy evolution of mean values. Taking the ECH results for granted 
this question can be
answered: The \msbar\ perturbative part of the power ansatz will 
fulfill the RGE.
To  leading order this is also true for the power contribution as its energy
dependence is similar to that of the coupling. As the size
of the power term is fit to the data the discrepancy to the ECH
expectation must be small. Still the predictability of the size of the power
terms (see Fig.~\ref{f:a0rgi}) using a single value of \asZ\ as input gives
preference to the ECH description for mean values.

It is likely that a large part of the annoying and unsatisfactorily
large spread of the \asZ\ results obtained from different event shape 
observables using
standard analyses is due to perturbing logarithmic terms induced by the choice
of the so-called physical scale and the \msbar\ 
renormalisation scheme. Given the above quoted positive ECH results 
experimentalists
should be encouraged to also analyse their data using the ECH scheme 
where possible and not 
regard this alternative to the  \msbar\ convention as a 
heretical digression. 

\begin{acknowledgments}
I would like to thank the organisers of the FRIF workshop, Gavin Salam, 
Mrinal Dasgupta and Yuri Dokshitzer for creating 
the pleasant
and truly ``workshoppish'' atmosphere at Jussieu and for the
possibility to 
summarise work prepared over the lifetime of \lep\ and \delphi.
I thank J.Drees and D.Wicke for comments to the manuscript.
\end{acknowledgments}
 

\end{document}